\documentclass[a4paper]{jpconf}
\usepackage{graphicx}
\usepackage{upgreek}
\usepackage{amsmath}
\usepackage{cases}

\begin{document}
\title{Absolute specific heat measurements of a microgram Pb crystal using ac nanocalorimetry}

\author{S Tagliati and A Rydh}

\address{Department of Physics, Stockholm University, AlbaNova, SE - 106 91 Stockholm, Sweden}

\ead{Stella.Tagliati@fysik.su.se}

\begin{abstract}
Heat capacity measurements using the ac steady state method are often considered difficult to provide absolute accuracy. By adjusting the working frequency to maintain a constant phase and using the phase information to obtain the heat capacity, we have found that it is possible to achieve good absolute accuracy.
Here we present a thermodynamic study of a $\sim2.6\,\upmu\mathrm{g}$ Pb superconducting crystal to demonstrate the newly opened capabilities. The sample is measured using a differential membrane-based calorimeter. The custom-made calorimetric cell is a pile of thin film Ti heater, insulation layer and $\mathrm{Ge}_\mathrm{1-x}\mathrm{Au}_\mathrm{x}$ thermometer fabricated in the center of two Si$_3$N$_4$ membranes. It has a background heat capacity $<100\,\mathrm{nJ/K}$ at $300\,\mathrm{K}$, decreasing to $9\,\mathrm{pJ/K}$ at $1\,\mathrm{K}$. 
The sample is characterized at temperatures down to $0.5\,\mathrm{K}$. The zero field transition at $T_\mathrm{c}=7.21\,\mathrm{K}$ has a width $\approx 20\,\mathrm{mK}$ and displays no upturn in $C$. From the heat capacity jump at $T_\mathrm{c}$ and the extrapolated Sommerfeld term we find $\Delta C/\gamma T_\mathrm{c}=2.68$. The latent heat curve obtained from the zero field heat capacity measurement, and the deviations of the thermodynamic critical field from the empirical expression $H_\mathrm{c}=H_\mathrm{c}(0)\left[1-\left(T/T_\mathrm{c}\right)^2\right]$ are discussed. Both analyses give results in good agreement with literature. 
\end{abstract}
\section{Introduction}
Calorimetry is a powerful tool that allows complete thermodynamic characterization of materials. In particular, nanocaloric measurements are suitable to study phase transitions and specific heat dependencies of new superconductors, often available in just $\upmu$g quantities. In the past two decades much attention was devoted to studies of high-$T_\mathrm{c}$ superconductors \cite{Fisher} and, more recently, the iron-based superconductors \cite{Paglione}. The electronic specific heat is one of the crucial parameters for understanding high-temperature superconductivity, but it is very hard to measure with good absolute accuracy. Furthermore, it represents just a fraction of the total heat capacity and therefore requires high resolution techniques. The AC steady state method \cite{SullivanSeidel} is a very sensitive technique which usually gives only relative heat capacity values \cite{Fisher} because of the difficulties related to the choice of the working frequency. 
\begin{figure}[ht]
\includegraphics[width=0.43\linewidth]{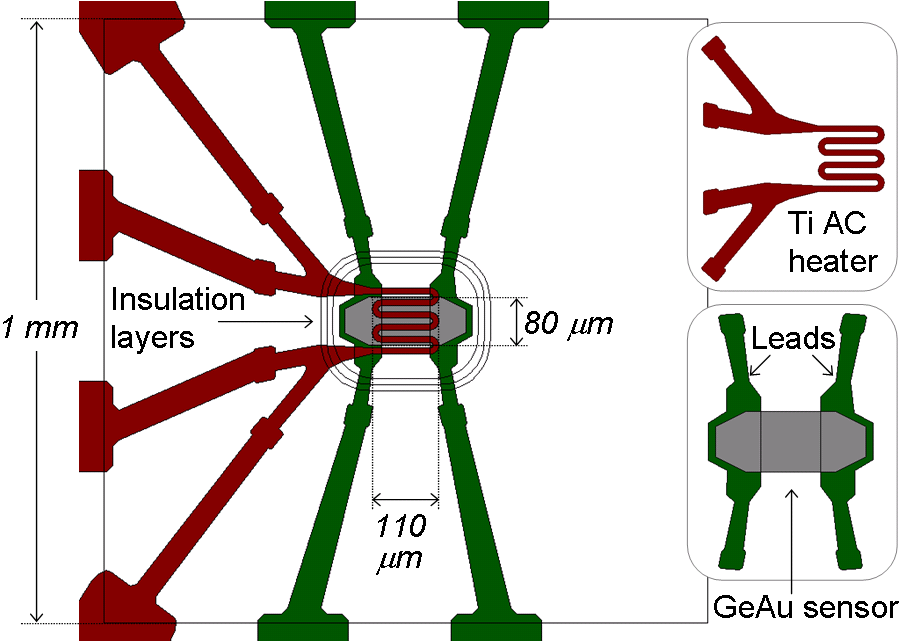}\hspace{0.04\linewidth}%
\begin{minipage}[b]{0.53\linewidth}\caption{\label{Fig1_Sketch}%
Layout of one of the $\mathrm{Si}_3\mathrm{N}_4$ membranes which host the calorimetric cell. A Ti heater and $\mathrm{Ge}_\mathrm{1-x}\mathrm{Au}_\mathrm{x}$ thermometer are fabricated in a pile in the central $110\times80\,\upmu\mathrm{m}^2$ area onto which the sample is placed. Two layers of $\mathrm{AlO}_\mathrm{x}$ ensure electrical insulation and a third $\mathrm{SiO}_2$ layer protects the surface of the $\mathrm{Ge}_\mathrm{1-x}\mathrm{Au}_\mathrm{x}$ sensor. The pile and the membrane underneath compose the platform with heat capacity $C_\mathrm{0}$. The membrane outside the platform thermally connects the sample area to the thermal bath (thermal conductance $K_\mathrm{e}$) and contributes with a heat capacity $C_{\mathrm{m,eff}}$.}
\end{minipage}
\end{figure}
In this method, a certain power $P_0\sin\omega t$ modulates the temperature of sample and calorimetric cell that oscillate with amplitude $T_{\mathrm{ac},0}$ and a phase lag $\phi$. The heat capacity is given by \cite{Gmelin}:
\begin{equation}{}\label{EqCK}
C =\frac{P_\mathrm{0}}{\omega T_\mathrm{ac,0}}\sin\phi.
\end{equation}
The optimal frequency range is usually quite narrow and depends on several factors, such as temperature, sample heat capacity and device thermal link. If $\omega$ is too high the sample becomes thermally disconnected and the signal probes just the heat capacity of the cell. On the other hand, the resolution degrades at too low frequencies \cite{Rydh_Manuscript}. 
We have implemented a measurement method which avoids these problems and furnishes absolute values \cite{Tagliati}. It requires continuous tuning of the frequency based on the reading of the phase.
The measured heat capacity can be written as:
\begin{equation}
C = C_0+C_\mathrm{m,eff} + (1-g)C_\mathrm{s},
\end{equation}\label{Eq_CKtotal}
where $C_0+C_\mathrm{m,eff}$ is the background heat capacity (see Fig.~\ref{Fig1_Sketch}) and $g$ is a frequency dependent function.
Inaccuracies in the evaluation of the sample heat capacity $C_\mathrm{s}$ are due to non-zero values of $g$ which, in turn, are caused by a finite thermal conductance $K_\mathrm{i}$ between sample and cell. The tangent of the phase $\tan\phi$ is an excellent indicator of the working conditions. It has a local maximum at a certain frequency $\omega_\mathrm{max}$, after which the absolute accuracy decreases quickly and the working conditions become unstable. The key parameter which controls this maximum is $\beta=K_\mathrm{i}/K_\mathrm{e}$: $\tan\phi_{\max}\approx\sqrt{\beta}/2$. In general, the absolute error at the maximum is $g|_{\omega_{\max}} \approx 1/\beta$. By tuning $\omega$ to keep a constant phase the error can be minimized while maintaining a good resolution. To demonstrate the effectiveness of the technique we here characterize a microgram Pb sample.
\section{Experimental}
The calorimetric cell used for measuring the Pb sample is shown in Fig.~\ref{Fig1_Sketch}. 
The crystal is selected to match the dimensions of the calorimeter central area. To reach a uniform temperature along the sample thickness at a given frequency, the crystal should be thinner than the length of the thermal wave. It was chosen to be less than $50\,\upmu\mathrm{m}$ thick. 
Apiezon grease, with thermal conductance $K_\mathrm{i}$, is used to ensure thermal contact between the sample and the pile of heater and thermometer underneath. Because of the design with all layers stacked on top of each other, the relaxation time within the calorimetric cell is negligible. 
In our case, $\beta \approx 700$ at room temperature. During each temperature scan the frequency is automatically varied to maintain the phase constant well below $\phi_\mathrm{max}$. Between $300\,\mathrm{K}$ and $0.5\,\mathrm{K}$ the drive frequency $f=\omega/4\pi$ spans the range $1\,\mathrm{Hz}<f<60\,\mathrm{Hz}$.
The temperature oscillation amplitude $T_{\mathrm{ac},0}$ is chosen to be a certain fraction of the absolute temperature $T$, and $P_0$ is changed accordingly. For a certain phase, the resolution improves with higher temperature oscillation amplitudes. For example, a typical resolution $\delta C/C=3\cdot10^{-4}$ is achieved at $T_{\mathrm{ac},0}/T=5\cdot10^{-3}$ for a $1\,\mathrm{s}$ integration time constant. The drawback of a high $T_{\mathrm{ac},0}/T$ ratio is a smearing effect of the absolute temperature scale. By decreasing $T_{\mathrm{ac},0}$ it is possible to contain this effect still maintaining a reasonable resolution. To measure the superconducting transition we used $T_{\mathrm{ac},0}=4.8\,\mathrm{mK}$ and $5\,\mathrm{s}$ time constant which gave $\delta C/C=6.8\cdot10^{-4}$.
The empty cell, and cell with grease were pre-characterized to be able to subtract each contribution from the total signal.
To explain the residual $\gamma$ in the superconducting state and obtain absolute agreement with the specific heat at $T_\mathrm{c}$, $c=1.09\,\mathrm{J}/{\mathrm{mol}\,\mathrm{K}}$ reported by Shiffman \textit{et al.} \cite{Schiffman}, $5\%$ of the normal state heat capacity had to be subtracted in addition to the device background. Whether this unexplained sample contribution is in reality due an absolute accuracy issue is left as an open question.
\section{Results}
\begin{figure}
\begin{center}
\includegraphics[height=12.2pc]{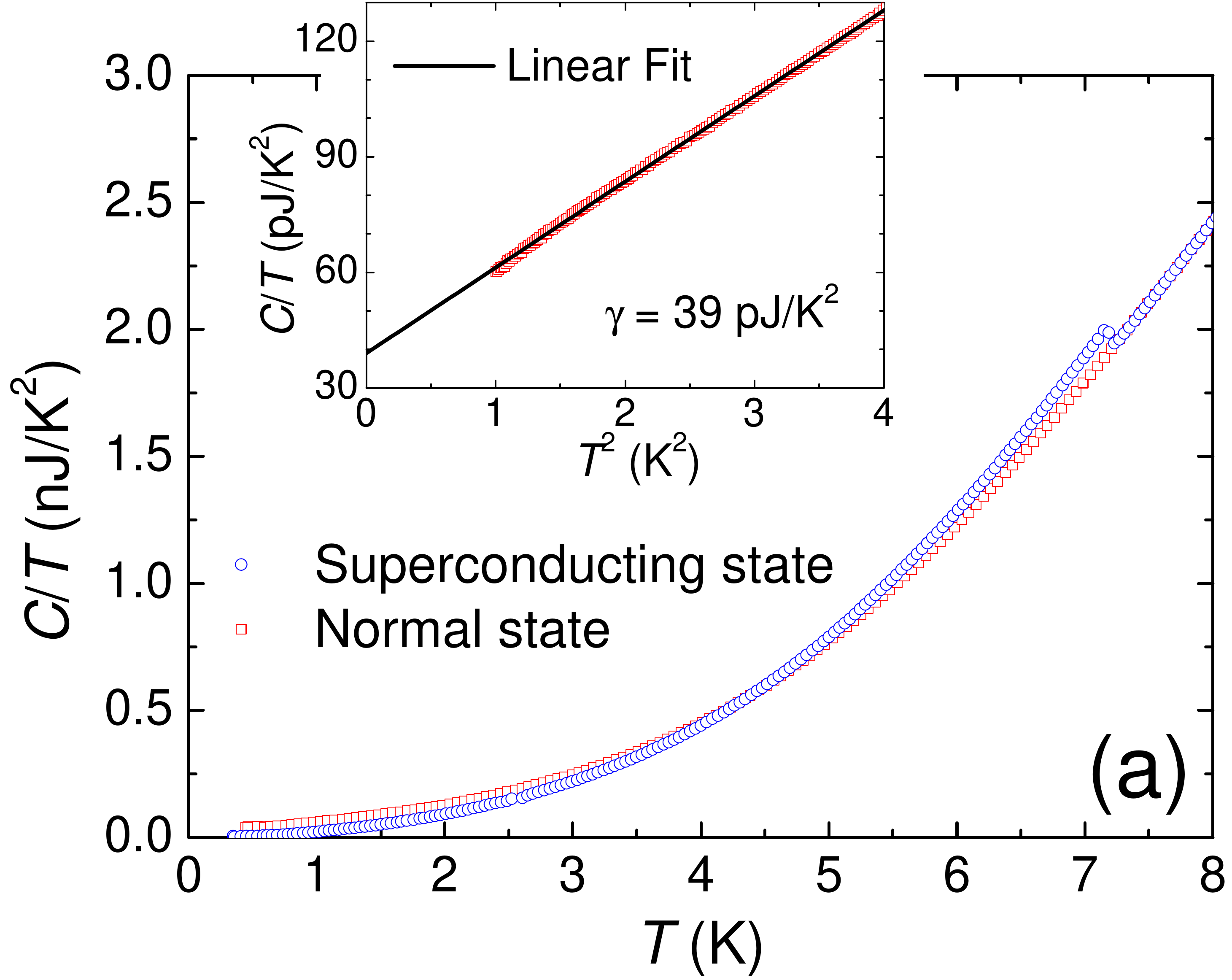}\hspace{2pc}%
\includegraphics[height=11.5pc]{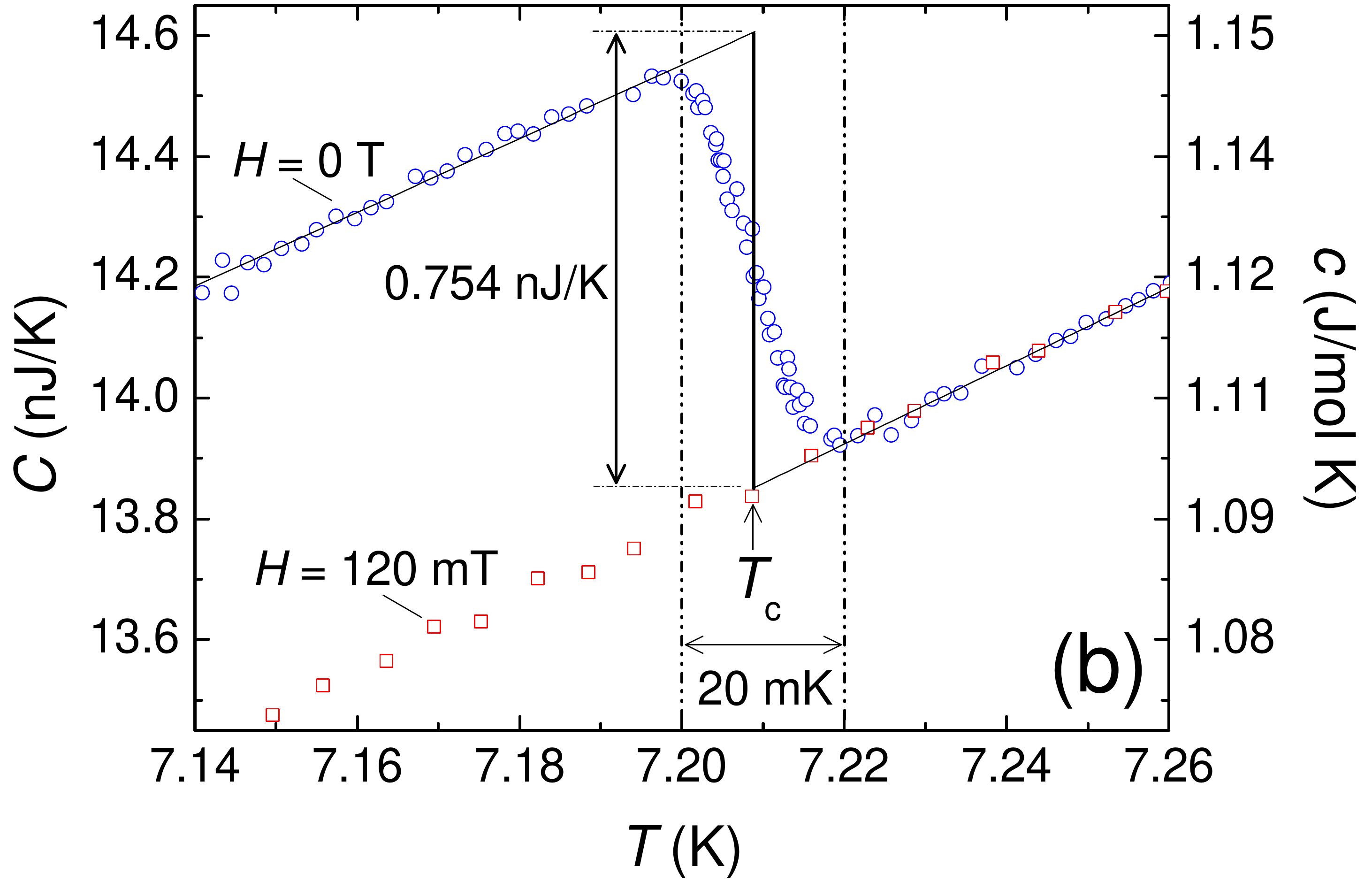}%
\end{center}
\caption{\label{Fig2_HeatCapacity}\textbf{(a)} Temperature dependence of $C/T$ of the Pb sample in the superconducting (circles) and normal (squares) states. The inset shows $C/T$ vs $T^2$ at low $T$ for the normal state. The line represent the best fit to $C/T=\gamma+\beta T^2$. \textbf{(b)} Heat capacity in zero field and in $120\,\mathrm{mT}$ around $T_\mathrm{c}$.}
\end{figure}

The Pb sample was characterized down to $0.5\,\mathrm{K}$. To obtain a pure superconducting state curve a small field was applied to compensate the remanent field of the magnet. The normal state was measured in $120\,\mathrm{mT}$. The curves $C(T)/T$ of the Pb sample are shown in Fig.~\ref{Fig2_HeatCapacity}a. The low temperature data (inset of Fig.~\ref{Fig2_HeatCapacity}a) were fitted by a linear function to extrapolate the normal state electronic heat capacity at zero temperature $\gamma=39.0\,\mathrm{pJ}/\mathrm{K}^2$. 
The zero field superconducting transition displayed in Fig.~\ref{Fig2_HeatCapacity}b, has a $20\,\mathrm{mK}$ width and no upturn in $C$. 
From the step height $\Delta C=0.754\,\mathrm{nJ}/\mathrm{K}$ at $T_\mathrm{c}$ and the Sommerfeld term $\gamma$, we find $\Delta C/\gamma T_\mathrm{c}=2.68$ in agreement with the values obtained from magnetic \cite{Decker} and other calorimetric \cite{Schiffman, Neighbor} measurements. The normalized slope of the specific heat discontinuity at $T_\mathrm{c}$ is an indicator of the coupling strength of the superconductor. From the data in Fig.~\ref{Fig2_HeatCapacity} we obtain $(T_\mathrm{c}/\Delta C)(d\Delta C/dT)_{T_\mathrm{c}}\approx4.5$, close to the value $4.6$ obtained from strong-coupling theory \cite{Carbotte}.

\begin{figure}
\begin{center}
\includegraphics[height=11.5pc]{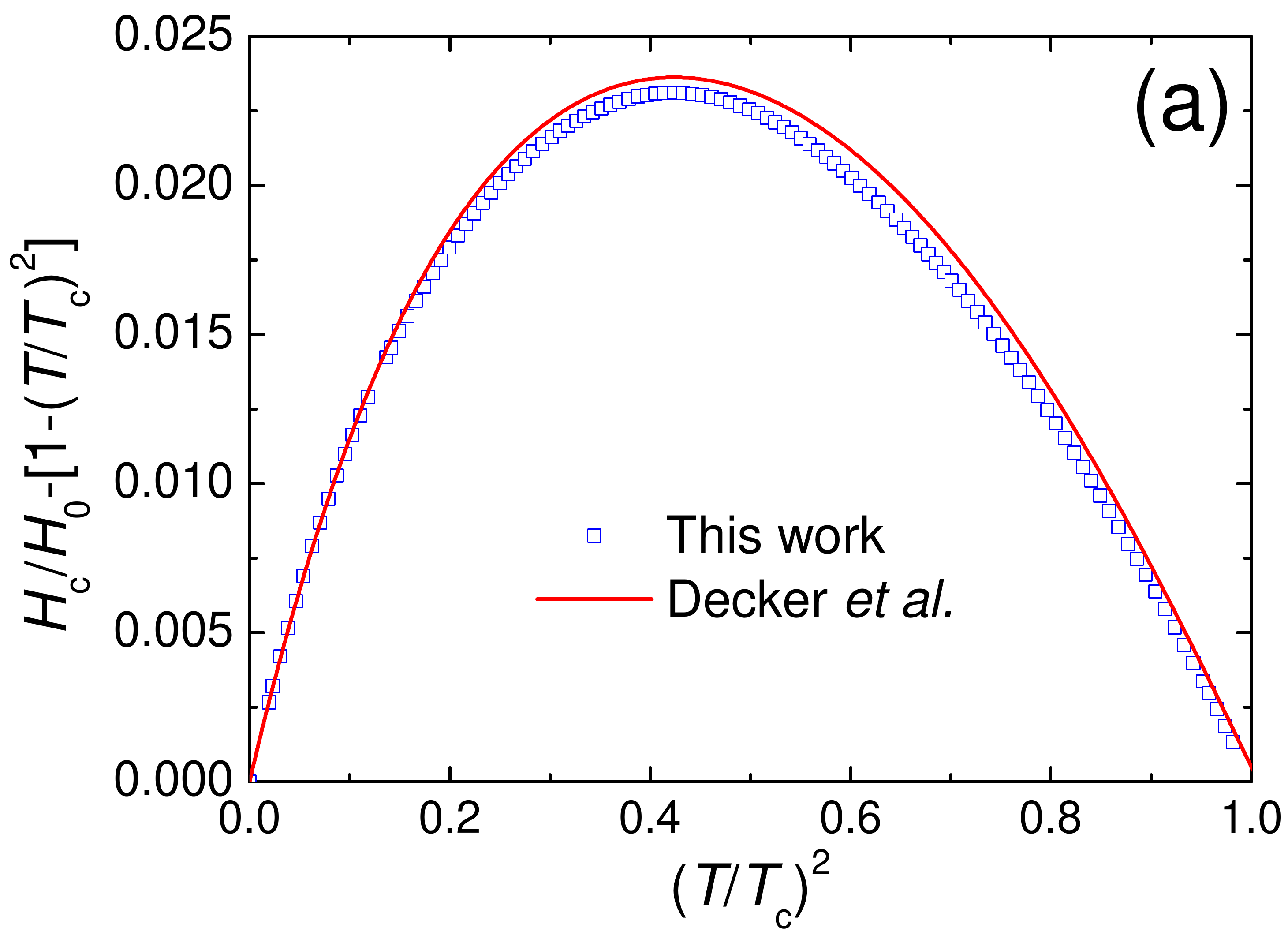}\hspace{2pc}%
\includegraphics[height=11.3pc]{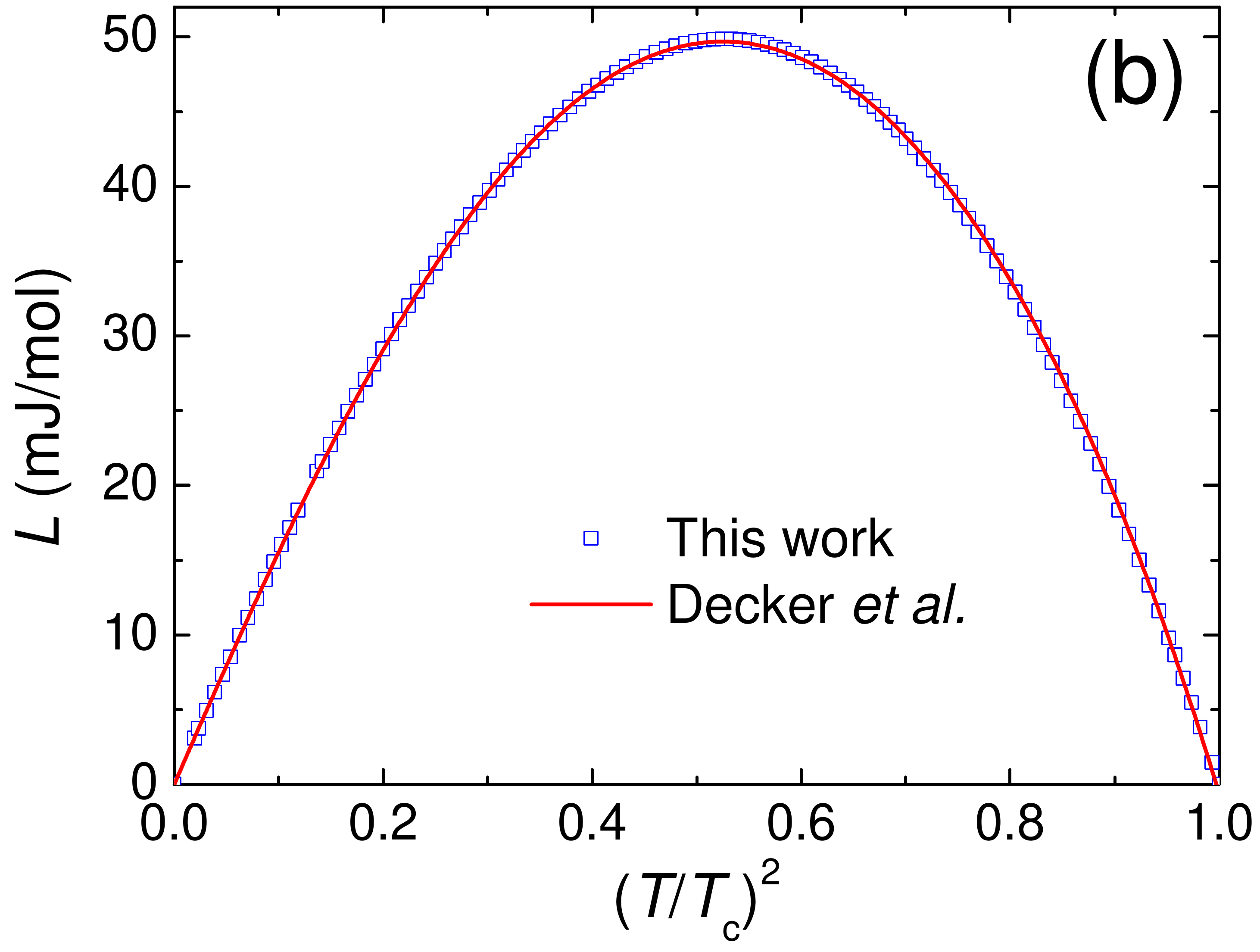}%
\end{center}
\caption{\label{Fig3_Hc&L}\textbf{(a)} Deviation of the reduced critical field from a parabolic curve as a function of the reduced temperature. \textbf{(b)} Transitional latent heat as a function of reduced temperature. In both figures the squares are values obtained in this experiment and the line is obtained from the analytical expression for $H_\mathrm{c}(T)$ (a) and $H_\mathrm{c}^2(T)$ (b) proposed by Decker \textit{et al.} \cite{Decker}.}
\end{figure}

\noindent Using the experimentally determined $\Delta C(T)$, one can calculate $\Delta S(T)$ and $\Delta U(T)$ by integrating $\Delta C(T)/T$ and $\Delta C(T)$ respectively, from $0\,\mathrm{K}$ to $T_\mathrm{c}$. By imposing the conservation law $\Delta F=F_\mathrm{s}-F_\mathrm{n}=0$ at $T_\mathrm{c}$ on the free energy difference $\Delta F=\Delta U-T\Delta S$, the condensation energy $\upmu_0H_\mathrm{c}^2 V/2$, is obtained.
To obtain the most precise measurement of the sample volume, $H_\mathrm{c}(0)$ was taken to be equal to the literature value $80.3\,\mathrm{mT}$ \cite{Carbotte}. The volume is equal to $2\Delta F (0)/\upmu_0 H^2_\mathrm{c}(0)$, giving $N=12.75\,\mathrm{nmol}$.
The temperature dependence of the deviation of $H_\mathrm{c}$ from the two fluid model, $D=H_\mathrm{c}(T)/H_\mathrm{c}(0)-[1-(T/T_\mathrm{c})^2]$ is plotted in Fig.~\ref{Fig3_Hc&L}a. Pb shows a positive deviation curve related to the strong electron-phonon coupling, in contrast with the negative $D$ values of the weak-coupling superconductors. The position of the maximum deviation is at $T/T_\mathrm{c}=0.65$, close to the experimentally determined functional form provided by Decker \textit{et al.} \cite{Decker}, the experimental value by Chanin \textit{et al.} \cite{Chanin}, and the strong-coupling theory calculation of Swihart \textit{et al.} \cite{Swihart}. The maximum deviation for our data, $2.3\%$, is slightly lower than the the values of Decker, $2.4\%$, and Swihart, $2.7\%$, but higher than the $2.1\%$ found by Chanin. 
The latent heat $L(T)=T\Delta S(T)$ is shown in Fig.~\ref{Fig3_Hc&L}b. Our data match the curve derived using the power series expansion of $H_\mathrm{c}^2(T)$ by Decker \textit{et al.} \cite{Decker}. The maximum falls at $T/T_\mathrm{c}=0.53$ for both curves.
\section{Summary and conclusions}
We recently proposed an experimental procedure to obtain both good resolution and absolute accuracy in AC calorimetry \cite{Tagliati}. The results reported in this paper on a $\sim2.6\,\upmu\mathrm{g}$ Pb sample demonstrate the feasibility and effectiveness of this method. From the presented low temperature superconducting and normal state heat capacity curves, thermodynamic properties such as Sommerfeld term $\gamma=3.06\,\mathrm{mJ/mol K}$, reduced jump anomaly $\Delta C/\gamma T_\mathrm{c}=2.68$ and normalized slope $(T_\mathrm{c}/\Delta C)(d\Delta C/dT)_{T_\mathrm{c}}\approx4.5$, are found in good agreement with literature values. The accuracy achieved is further confirmed by the obtained temperature dependence of $H_\mathrm{c}(T)$ and latent heat.
\ack
Support from the the Swedish Research Council and the Knut and Alice Wallenberg Foundation is acknowledged. We would like to thank V.~M.~Krasnov for useful discussions and for providing the Pb sample.


\end{document}